\documentclass[12pt, reqno]{amsart}
\usepackage[margin=1in]{geometry}
\usepackage{amssymb,latexsym,amsmath,amscd,amsfonts}
\usepackage{latexsym}
\usepackage[mathscr]{eucal}
\usepackage{bm}
\usepackage{mathptmx}
\usepackage{amssymb}
\usepackage{amsthm}
\usepackage{dcolumn}
\usepackage[all]{xy}
\usepackage{enumitem}
\usepackage[utf8]{inputenc}

\def\textmatrix#1&#2\\#3&#4\\{\bigl({#1 \atop #3}\ {#2 \atop #4}\bigr)}
\def\dispmatrix#1&#2\\#3&#4\\{\left({#1 \atop #3}\ {#2 \atop #4}\right)}
\newcommand{\beg}{\begin{equation}}
	\newcommand{\eeg}{\end{equation}}
\newcommand{\ben}{\begin{eqnarray*}}
	\newcommand{\een}{\end{eqnarray*}}

\numberwithin{equation}{section} \theoremstyle{definition}

\def\textmatrix#1&#2\\#3&#4\\{\bigl({#1 \atop #3}\ {#2 \atop #4}\bigr)}
\def\dispmatrix#1&#2\\#3&#4\\{\left({#1 \atop #3}\ {#2 \atop #4}\right)}

\newtheorem{theorem}{Theorem}[section]
\newtheorem{lemma}[theorem]{Lemma}

\theoremstyle{definition}

\theoremstyle{remark}
\newtheorem{remark}[theorem]{Remark}

\numberwithin{equation}{section}

\usepackage{color}
\usepackage{graphicx}
\usepackage{subcaption}

\newcommand{\eps}{\epsilon}
\newcommand{\vecone}{\boldsymbol{1}}
\newcommand{\identity}{\boldsymbol{I}}
\newcommand{\zero}{\boldsymbol{{0}}}
\newcommand{\matone}{\boldsymbol{J}}
\newcommand{\hatmat}{\boldsymbol{H}}

\begin{document}
	\title[Studying Optimal Designs for Multivariate Crossover Trials]{Studying Optimal Designs for Multivariate Crossover Trials}
	\thanks{This is a preprint of an article that has been accepted for publication in Contemporary Mathematics (American Mathematical Society).}
\author{Shubham Niphadkar}
\address{Department of Mathematics, Indian Institute of Technology Bombay, Powai, Mumbai, Maharashtra 400076, India}
\curraddr{Department of Mathematics, Indian Institute of Technology Bombay, Powai, Mumbai, Maharashtra 400076, India}
\email{niphadkarshubham@gmail.com}

\author{Siuli Mukhopadhyay}
\address{Department of Mathematics, Indian Institute of Technology Bombay, Powai, Mumbai, Maharashtra 400076, India}
\email{siuli@math.iitb.ac.in}


	
	
	\begin{abstract}
		This article discusses $A$-, $D$- and $E$-optimality results for multivariate crossover designs, where more than one response is measured from every period for each subject. The motivation for these multivariate designs comes from a $3 \times 3$ crossover trial that investigates how an oral drug affects biomarkers of mucosal inflammation, by analyzing the various gene profiles from each participant. A multivariate response crossover model with fixed effects including direct and carryover effects, and with heteroscedastic error terms is considered to fit the multiple responses measured. It is assumed all throughout the article that there is no correlation between responses but there is presence of correlation within responses. Corresponding to the direct effects, we obtain the information matrix in a multiple response setup. Various results regarding this information matrix are studied. For $p$ periods and $t$ treatments, orthogonal array design of type $I$ and strength $2$ is proved as $A$-, $D$- and $E$-optimal, when $p=t \geq 3$. 
	\end{abstract}

	\maketitle
	
\section{Introduction}
In many clinical studies, crossover designs with measurements on multiple responses from each subject in each period are utilized. \cite{7} and \cite{22} used a multivariate crossover design to measure systolic and diastolic blood pressure, and to measure multiple gene expression profiles, respectively. However in these studies no investigation of optimality of such multivariate crossover designs were done. The researchers instead considered various structures for between and within correlation in responses, and discussed methods of estimation in such crossover trials. We explore optimality of exact multiple response crossover designs, in this study. For a given design, by denoting the number of treatments and periods by $t$ and $p$, respectively, we obtain $A$-, $D$- and $E$-optimal designs for the case when $p=t \geq 3$.\par
Crossover designs find application in various areas like pharmaceutical studies, clinical trials, agricultural sciences and biological studies (\cite{5, 28}). Various researchers including \cite{8}, \cite{9}, \cite{4}, \cite{16}, \cite{30}, \cite{18}, \cite{19}, \cite{17}, \cite{10}, \cite{11} and \cite{27} have studied optimality of univariate crossover designs considering a fixed effect model, while \cite{20}, \cite{21}, \cite{2}, \cite{3} and \cite{12} considered a random subject effect model. Optimal crossover designs, where responses are non-normal in nature were explored by \cite{28}, \cite{13}, \cite{24} and \cite{29}, to name a few. However, all these works are based on single response crossover trials. Recently, \cite{25} investigated universal optimality of multivariate crossover design, where the responses are uncorrelated and have same within response correlation structure, whereas \cite{26} studied efficient designs when the responses are correlated, under the proportional and the markovian covariance structure.\par
In this chapter, we introduce a multivariate response model with fixed effects that incorporates both direct as well as carryover effects of treatments for modelling data from a multivariate crossover trial. We consider the responses to be uncorrelated, while each response is considered to be correlated within themselves with different responses having different correlation structure and heteroscedastic variances. In this case, we prove that corresponding to direct effects, the information matrix is not completely symmetric. Therefore, to find $A$-, $D$- and $E$-optimal designs we resort to eigenvalue analysis of the information matrix.

\section{Proposed statistical model}
\label{models}
We examine $p \times t$ crossover designs where $t \geq 2$. Here $t$ represent treatments which are allocated over $p$ periods and $n$ experimental subjects. We denote the collection of all such crossover designs by $\Omega_{t,n,p}$. In this work, we consider that corresponding to a design $d \in \Omega_{t,n,p}$, for every subject in each period, the observations are recorded on $g$ response variables, where $g \geq 1$. For the observation measured in the $i^{th}$ period and $j^{th}$ subject on the $k^{th}$ response, the corresponding random variable $Y_{dijk}$ is modelled as:
\begin{equation}
Y_{dijk} = \mu_k + \alpha_{i,k} + \beta_{j,k} + \tau_{d \left( i,j \right),k} + \rho_{d \left( i-1,j \right),k}  + \eps_{ijk},
\label{model1}
\end{equation}
where $1 \leq i \leq p$, $1 \leq j \leq n$ and $1 \leq k \leq g$. Here $\mu_k$, $\alpha_{i,k}$,  $\beta_{j,k}$ denote the intercept, effect due to the $i^{th}$ period and effect due to the $j^{th}$ subject, respectively, on the $k^{th}$ response. The direct and the carryover effect of first order on the $k^{th}$ response due to treatment $s$ is denoted as $\tau_{s,k}$ and $\rho_{s,k}$, respectively, where $1 \leq s \leq t$. We assume all the effects to be fixed effects. The corresponding error term is denoted by $\eps_{ijk}$, where we assume that $\mathbb{E} \left( \eps_{ijk} \right) = 0$, $\mathbb{\text{Cov}} \left( \eps_{ijk}, \eps_{i'jk'}\right)=0$, for $1 \leq i,i' \leq p$ and $1 \leq k \neq k' \leq g$, and $\mathbb{\text{Cov}} \left( \eps_{ijk}, \eps_{i'j'k'}\right)=0$, for $1 \leq i,i' \leq p$, $1 \leq j \neq j' \leq n$ and $1 \leq k,k' \leq g$. These imply that there is no correlation between the observations taken on distinct response variables. Also as in most crossover trials, we assume that the observations from distinct subjects are uncorrelated. In the above model, we have considered that the intercept, effects due to periods, subjects, treatments and first order carryover effects change according to the response variable. We assume that there is no carryover effect associated with the first period, thus $\rho_{d \left( 0 , j \right),k} =0$.\par
Next, we discuss steps to write model \eqref{model1} in matrix form. For $d \in \Omega_{t,n,p}$, we let $\boldsymbol{Y}_{dk} = \left( Y_{d11k}, \cdots, Y_{dp1k}, Y_{d12k}, \cdots, Y_{dp2k}, \cdots, Y_{d1nk}, \cdots, Y_{dpnk} \right)^{'}$, $\boldsymbol{\eps}_k = \\ \left( \eps_{11k}, \cdots, \eps_{p1k}, \eps_{12k}, \cdots, \eps_{p2k}, \cdots, \eps_{1nk}, \cdots, \eps_{pnk} \right)^{'}$, $\boldsymbol{\alpha}_k = \left( \alpha_{i,k} \right)_{i=1}^{p}$, $\boldsymbol{\beta}_k = \left( \beta_{j,k} \right)_{j=1}^{n}$, $\boldsymbol{\tau}_k = \left( \tau_{s,k} \right)_{s=1}^{t}$ and $\boldsymbol{\rho}_k = \left( \rho_{s,k} \right)_{s=1}^{t}$ which are the column vectors. Here for the $k^{th}$ response, $\boldsymbol{\alpha}_k$, $\boldsymbol{\beta}_k$, $\boldsymbol{\tau}_k$ and $\boldsymbol{\rho}_k$ are the vectors of period, subject, direct and carryover effects, respectively. Note that in the $g$ variable case, we have $g$ number of $\boldsymbol{\tau}_k$'s and $\boldsymbol{\rho}_k$'s. The parameter vector is denoted by $\boldsymbol{\theta}_k = \left( \mu_k, \boldsymbol{\alpha}_k^{'}, \boldsymbol{\beta}_k^{'}, \boldsymbol{\tau}_k^{'}, \boldsymbol{\rho}_k^{'} \right)^{'}$.\par
The incidence matrix of the period versus direct effects for the $j^{th}$ subject, $\boldsymbol{T}_{dj}$, is a $p \times t$ matrix, while the incidence matrix of the period versus carryover effects, $\boldsymbol{F}_{dj}$, is also of dimension $p \times t$. For each $j$, $\boldsymbol{F}_{dj}$ can be written as $\boldsymbol{F}_{dj} =
\begin{bmatrix}
\zero^{'}_{p-1 \times 1} & 0\\
\identity_{p-1} & \zero_{p-1 \times 1}
\end{bmatrix}
\boldsymbol{T}_{dj}
$. Defining $\boldsymbol{X}_{dk}=
\begin{bmatrix}
\vecone_{np} & \boldsymbol{P} & \boldsymbol{U} & \boldsymbol{T}_d & \boldsymbol{F}_d
\end{bmatrix}$, model \eqref{model1} can be represented in matrix form as
\begin{equation}
\boldsymbol{Y}_{dk} = \boldsymbol{X}_{dk} \boldsymbol{\theta}_k + \boldsymbol{\eps}_k,
\label{model11}
\end{equation}
where 
$\boldsymbol{T}_d = \begin{bmatrix}
\boldsymbol{T}^{'}_{d1} &
\cdots &
\boldsymbol{T}^{'}_{dn}
\end{bmatrix}^{'}
$, $
\boldsymbol{F}_d =
\begin{bmatrix}
\boldsymbol{F}^{'}_{d1} &
\cdots &
\boldsymbol{F}^{'}_{dn}
\end{bmatrix}^{'}
$, $\boldsymbol{P}=\vecone_n \otimes \identity_p$ and $\boldsymbol{U}=\identity_n \otimes \vecone_p$. Using $\boldsymbol{X_1} = 
\begin{bmatrix} 
\boldsymbol{P} & \boldsymbol{U}
\end{bmatrix}$ and $\boldsymbol{X_2} = 
\begin{bmatrix}  
\boldsymbol{T}_d & \boldsymbol{F}_d 
\end{bmatrix}$, $\boldsymbol{X}_{dk}$ is written as
\begin{equation}
\boldsymbol{X}_{dk} =
\begin{bmatrix}
\vecone_{np} & \boldsymbol{X_1} & \boldsymbol{X_2}
\end{bmatrix}.
\label{Xd}
\end{equation}
\par
Combining the $g$ models in \eqref{model11}, we get
\begin{equation}
\begin{split}
\begin{bmatrix}
 \boldsymbol{Y}^{'}_{d1} &
\cdots &
\boldsymbol{Y}^{'}_{dg}
\end{bmatrix}^{'}
&= \left( \oplus_{k=1}^{g} \boldsymbol{X}_{d1} \right)
\begin{bmatrix}
\boldsymbol{\theta}^{'}_1 &
\cdots &
\boldsymbol{\theta}^{'}_g
\end{bmatrix}^{'}
+
\begin{bmatrix}
 \boldsymbol{\eps}^{'}_{1} &
\cdots &
\boldsymbol{\eps}^{'}_{g}
\end{bmatrix}^{'} \\
&= \left( \identity_{g} \otimes \boldsymbol{X}_{d1} \right) 
\begin{bmatrix}
\boldsymbol{\theta}^{'}_1 &
\cdots &
\boldsymbol{\theta}^{'}_g
\end{bmatrix}^{'}
+
\begin{bmatrix}
 \boldsymbol{\eps}^{'}_{1} &
\cdots &
\boldsymbol{\eps}^{'}_{g}
\end{bmatrix}^{'}.
\end{split}
\label{unio}
\end{equation}
Along with the assumptions of zero mean errors and uncorrelated response variables, we consider that the dispersion matrix of $\boldsymbol{\eps}_k$ is given as $\mathbb{D} \left( \boldsymbol{\eps}_k \right) = \sigma_k^2 \boldsymbol{\Sigma}_k = \sigma_k^2 \left( \identity_{n} \otimes \boldsymbol{V}_k \right)$, where $\boldsymbol{V}_k$ is a $p \times p$ known positive definite and symmetric matrix, and $\sigma_k^2 >0$ is an unknown constant. We assume that $\sigma_k^2$'s and $\boldsymbol{V}_k$'s are not the same for different responses.

\section{Information matrix}
\label{information matrices}
In this section, corresponding to the direct effects we derive the information matrix, when $g > 1$. For $g=1$, we have the univariate response for which the information matrix is studied in \cite{1} for the model \eqref{unio}. For $d \in \Omega_{t,n,p}$, corresponding to the direct effects let the information matrix be denoted by $\boldsymbol{C}_{d}$.\par
Note that here we consider within subject correlation in the errors. The dispersion matrix of $\boldsymbol{\eps} = \begin{bmatrix}
\boldsymbol{\eps}^{'}_1 &
\cdots &
\boldsymbol{\eps}^{'}_g
\end{bmatrix}^{'}$ is
\begin{equation}
\mathbb{D} \left( 
\boldsymbol{\eps}
\right)
= \boldsymbol{\Sigma} =
\sigma_1^2 \boldsymbol{\Sigma}_1 \oplus \cdots \oplus \sigma_g^2 \boldsymbol{\Sigma}_g = \oplus_{k=1}^{g} \sigma_k^2 \boldsymbol{\Sigma}_k,
\label{dispepsc}
\end{equation}
where $\boldsymbol{\Sigma}_k = \identity_n \otimes \boldsymbol{V}_k$, and $\sigma_k^2$ and $\boldsymbol{V}_k$ vary with $k$.\par
\begin{lemma}
The information matrix for direct effects is given by
\begin{equation}
\boldsymbol{C}_{d} = \sigma_1^{-2} \boldsymbol{C}_{d(1)} \oplus \cdots \oplus \sigma_g^{-2} \boldsymbol{C}_{d(g)}, \label{x4}
\end{equation}
where
\begin{equation}
\begin{split}
\boldsymbol{C}_{d(k)} &= \boldsymbol{C}_{{d(k)}11} - \boldsymbol{C}_{{d(k)}12} \boldsymbol{C}^{-}_{{d(k)}22} \boldsymbol{C}_{{d(k)}21}\\
&=\boldsymbol{T}^{'}_d  \boldsymbol{\Sigma}_k^{-1/2} \text{\rmfamily\upshape pr}^{\perp} \left( \boldsymbol{\Sigma}_k^{-1/2} 
\begin{bmatrix}
\boldsymbol{X_1} &  \boldsymbol{F}_d
\end{bmatrix}  
\right) \boldsymbol{\Sigma}_k^{-1/2} \boldsymbol{T}_d. \label{x7}
\end{split}
\end{equation} 
Here $\boldsymbol{C}_{{d(k)}11}=\boldsymbol{T}^{'}_d \boldsymbol{A}^{*}_k \boldsymbol{T}_d$, $\boldsymbol{C}_{{d(k)}12}=\boldsymbol{C}_{{d(k)}21}^{'}=\boldsymbol{T}^{'}_d \boldsymbol{A}^{*}_k \boldsymbol{F}_d$, $\boldsymbol{C}_{{d(k)}22}=\boldsymbol{F}^{'}_d \boldsymbol{A}^{*}_k \boldsymbol{F}_d$, $\boldsymbol{A}^{*}_k = \hatmat_n \otimes \boldsymbol{V}^{*}_k$, $\boldsymbol{V}^{*}_k = \boldsymbol{V}_k^{-1} - \delta_k \boldsymbol{V}_k^{-1} \matone_{p \times p} \boldsymbol{V}_k^{-1}$ and $\delta_k =\left( \vecone^{'}_p \boldsymbol{V}_k^{-1} \vecone_p \right)^{-1}$.
\label{thm4c}
\end{lemma}

\begin{proof}
\setcounter{MaxMatrixCols}{20}
Let $\boldsymbol{\xi}_k=\begin{bmatrix} \mu_k & \boldsymbol{\alpha}^{'}_k &   \boldsymbol{\beta}^{'}_k & \boldsymbol{\rho}^{'}_k \end{bmatrix}^{'}$, $\boldsymbol{\eta}^{(1)} = \begin{bmatrix}
\boldsymbol{\tau}^{'}_1 & \cdots & \boldsymbol{\tau}^{'}_g
\end{bmatrix}^{'}$ and $\boldsymbol{\eta}^{(2)} = \begin{bmatrix}
\boldsymbol{\xi}_1^{'} & \cdots & \boldsymbol{\xi}_g^{'}
\end{bmatrix}^{'}
$. Then by rearranging the parameters, we represent the model \eqref{unio} as
\begin{multline}
\begin{bmatrix}
 \boldsymbol{Y}^{'}_{d1} &
 \cdots &
\boldsymbol{Y}^{'}_{dg}
\end{bmatrix}^{'}
= 
\left( \identity_g \otimes \boldsymbol{T}_d \right) \boldsymbol{\eta}^{(1)} + \left(  \identity_g \otimes
\begin{bmatrix}
 \vecone_{np}  & \boldsymbol{X_1} & \boldsymbol{F}_d
\end{bmatrix}
\right)
\boldsymbol{\eta}^{(2)} + \\ 
\begin{bmatrix}
 \boldsymbol{\eps}^{'}_{1} &
 \cdots &
\boldsymbol{\eps}^{'}_{g}
\end{bmatrix}^{'}.
\label{alt-model}
\end{multline}
Premultiplying \eqref{alt-model} by $\boldsymbol{\Sigma}^{-1/2}$, the model can be expressed as
\begin{multline}
\begin{bmatrix}
 \boldsymbol{Y}^{'}_{d1(new)} &
 \cdots &
\boldsymbol{Y}^{'}_{dg(new)}
\end{bmatrix}^{'}
= 
\boldsymbol{\Sigma}^{-1/2}
\left( \identity_g \otimes \boldsymbol{T}_d \right) \boldsymbol{\eta}^{(1)} + \\
\boldsymbol{\Sigma}^{-1/2} \left( \identity_g \otimes 
\begin{bmatrix}
 \vecone_{np}  & \boldsymbol{X_1} & \boldsymbol{F}_d
\end{bmatrix}
\right)
\boldsymbol{\eta}^{(2)} + 
\begin{bmatrix}
 \boldsymbol{\eps}^{'}_{1(new)} &
 \cdots &
\boldsymbol{\eps}^{'}_{g(new)}
\end{bmatrix}^{'},
\label{alt-model-1}
\end{multline}
where $\boldsymbol{Y}_{dk(new)} = \sigma_k^{-1} \boldsymbol{\Sigma}_k^{-1/2} \boldsymbol{Y}_{dk}$ and $\boldsymbol{\eps}_{k(new)} = \sigma_k^{-1} \boldsymbol{\Sigma}_k^{-1/2} \boldsymbol{\eps}_{k}$.\par
Therefore from \cite{15}, employing the expression for the information matrix, we have
\begin{equation*}
\boldsymbol{C}_{d} =
\left( \identity_g \otimes \boldsymbol{T}_d^{'} \right) \boldsymbol{\Sigma}^{-1/2} \text{\rmfamily\upshape pr}^{\perp} \left( \boldsymbol{\Sigma}^{-1/2}  \left( \identity_g \otimes 
\begin{bmatrix}
\vecone_{np} & \boldsymbol{X_1} & \boldsymbol{F}_d
\end{bmatrix}  
\right)
\right) \boldsymbol{\Sigma}^{-1/2} \left( \identity_g \otimes \boldsymbol{T}_d\right).
\end{equation*}
From \cite{1}, since $\vecone_{np}$ belongs to the column space of $\boldsymbol{X_1}$, we get
\begin{equation}
\boldsymbol{C}_{d} =
\left( \identity_g \otimes \boldsymbol{T}_d^{'} \right) \boldsymbol{\Sigma}^{-1/2} \text{\rmfamily\upshape pr}^{\perp} \left( \boldsymbol{\Sigma}^{-1/2}  \left( \identity_g \otimes 
\begin{bmatrix}
\boldsymbol{X_1} & \boldsymbol{F}_d
\end{bmatrix}  
\right)
\right) \boldsymbol{\Sigma}^{-1/2} \left( \identity_g \otimes \boldsymbol{T}_d\right).
\label{info-1}
\end{equation}
Using \eqref{dispepsc} in the above equation, we get
\begin{equation}
\begin{split}
\boldsymbol{C}_{d} &= \oplus_{k=1}^{g} \sigma_k^{-2} \boldsymbol{T}_d^{'} \boldsymbol{\Sigma}_k^{-1/2} \text{\rmfamily\upshape pr}^{\perp} \left( \boldsymbol{\Sigma}_k^{-1/2} \begin{bmatrix}
\boldsymbol{X_1} & \boldsymbol{F}_d
\end{bmatrix}   
\right) \boldsymbol{\Sigma}_k^{-1/2} \boldsymbol{T}_d \\
&= \sigma_1^{-2} \boldsymbol{C}_{d(1)} \oplus \cdots \oplus \sigma_g^{-2} \boldsymbol{C}_{d(g)}, \label{info-2}
\end{split}
\end{equation}
where 
\begin{equation}
\boldsymbol{C}_{d(k)} = \boldsymbol{T}_d^{'} \boldsymbol{\Sigma}_k^{-1/2} \text{\rmfamily\upshape pr}^{\perp} \left( \boldsymbol{\Sigma}_k^{-1/2} \begin{bmatrix}
\boldsymbol{X_1} & \boldsymbol{F}_d
\end{bmatrix}   
\right) \boldsymbol{\Sigma}_k^{-1/2} \boldsymbol{T}_d.
\label{info-3}
\end{equation}
Here $\boldsymbol{\Sigma}_k = \identity_n \otimes \boldsymbol{V}_k$, where $\boldsymbol{V}_k$ is a known positive definite and symmetric $p \times p$ matrix. So from \cite{1}, we get
\begin{equation}
\boldsymbol{T}_d^{'} \boldsymbol{\Sigma}_k^{-1/2} \text{\rmfamily\upshape pr}^{\perp} \left( \boldsymbol{\Sigma}_k^{-1/2} \begin{bmatrix}
\boldsymbol{X_1} & \boldsymbol{F}_d
\end{bmatrix}   
\right) \boldsymbol{\Sigma}_k^{-1/2} \boldsymbol{T}_d = 
\boldsymbol{C}_{{d(k)}11} - \boldsymbol{C}_{{d(k)}12} \boldsymbol{C}^{-}_{{d(k)}22} \boldsymbol{C}_{{d(k)}21},
\label{info-4}
\end{equation}
where $\boldsymbol{C}_{{d(k)}11}=\boldsymbol{T}^{'}_d \boldsymbol{A}^{*}_k \boldsymbol{T}_d$, $\boldsymbol{C}_{{d(k)}12}=\boldsymbol{C}_{{d(k)}21}^{'}=\boldsymbol{T}^{'}_d \boldsymbol{A}^{*}_k \boldsymbol{F}_d$, $\boldsymbol{C}_{{d(k)}22}=\boldsymbol{F}^{'}_d \boldsymbol{A}^{*}_k \boldsymbol{F}_d$, $\boldsymbol{A}^{*}_k = \hatmat_n \otimes \boldsymbol{V}^{*}_k$, $\boldsymbol{V}^{*}_k = \boldsymbol{V}_k^{-1} - \delta_k \boldsymbol{V}_k^{-1} \matone_{p \times p} \boldsymbol{V}_k^{-1}$ and $\delta_k =\left( \vecone^{'}_p \boldsymbol{V}_k^{-1} \vecone_p \right)^{-1}$.\par
Thus from \eqref{info-3} and \eqref{info-4}, we can prove \eqref{x7}.
\end{proof}

\begin{remark}
The information matrix $\boldsymbol{C}_{d}$ is symmetric, non-negative definite (n.n.d.) matrix having zero column sums and row sums. Moreover, $\boldsymbol{C}_{d}$ is invariant with respect to the choice of generalized inverses involved.
\label{re1c}
\end{remark}

\begin{proof}
Evidently, we have by \eqref{x4} and \eqref{x7} that $\boldsymbol{C}_{d}$ is symmetric and n.n.d. From Lemma~\ref{thm4c}, we get
\begin{equation}
\boldsymbol{C}_{d} = \sigma_1^{-2} \boldsymbol{C}_{d(1)} \oplus \cdots \oplus \sigma_g^{-2} \boldsymbol{C}_{d(g)}, 
\label{x4-1}
\end{equation}
where
\begin{equation*}
\boldsymbol{C}_{d(k)} 
=\boldsymbol{T}^{'}_d  \boldsymbol{\Sigma}_k^{-1/2} \text{\rmfamily\upshape pr}^{\perp} \left( \boldsymbol{\Sigma}_k^{-1/2} 
\begin{bmatrix}
\boldsymbol{X_1} &  \boldsymbol{F}_d
\end{bmatrix}  
\right) \boldsymbol{\Sigma}_k^{-1/2} \boldsymbol{T}_d.
\end{equation*} 
Here $\boldsymbol{\Sigma}_k = \identity_n \otimes \boldsymbol{V}_k$, where $\boldsymbol{V}_k$ is a $p \times p$ known positive definite and symmetric matrix. Hence from \cite{1}, we get that $\boldsymbol{C}_{d(k)} \vecone_t = \zero_{t \times 1}$, $\vecone_t^{'} \boldsymbol{C}_{d(k)}  = \zero_{1 \times t}$ and $\boldsymbol{C}_{d(k)}$ is invariant with respect to the choice of generalized inverses involved. Thus from \eqref{x4-1}, we get
\begin{equation*}
\begin{split}
\boldsymbol{C}_{d} \vecone_{gt} &= \oplus_{k=1}^{g} \sigma_k^{-2} \boldsymbol{C}_{d(k)} \vecone_t = \zero_{gt \times 1},\\
\vecone_{gt}^{'} \boldsymbol{C}_{d}  &= \oplus_{k=1}^{g} \sigma_k^{-2} \vecone_t^{'} \boldsymbol{C}_{d(k)}  = \zero_{1 \times gt}, \text{ and}
\end{split}
\end{equation*}
$\boldsymbol{C}_{d}$ is invariant with respect to the choice of generalized inverses involved. 
\end{proof}

\section{Optimality results}
\label{optimality-m}
Various authors including \cite{14}, \cite{9}, \cite{31}, \cite{6} and \cite{1} have discussed $A$-, $D$-, $E$- and universal optimality criterion for the $g=1$ setup. For the univariate response case, \cite{14} and \cite{31} have provided sufficient conditions for universal optimality. In this section, we extend these sufficient conditions for the multivariate response setup. Let for $g > 1$, $\mathcal{B}_{gt}$ be a class of $gt \times gt$ symmetric, non-negative definite (n.n.d.) matrices having zero row sums. Note that for the multivariate response case, the information matrix for the direct effects, $\boldsymbol{C}_{d}$, is a $gt \times gt$ matrix. From Remark~\ref{re1c}, it is clear that for $g>1$, $\boldsymbol{C}_{d} \in \mathcal{B}_{gt}$. In the next lemmas, we present sufficient conditions for obtaining universally optimal designs in the $g > 1$ setup. 

\begin{lemma}
Let a design $d^* \in \mathcal{D}$, where $\mathcal{D}$ is a subclass of designs, be such that under model \eqref{unio}, the information matrix $\boldsymbol{C}_{d^*}$ is completely symmetric and $d^*$ maximizes $\text{\rmfamily\upshape tr} \left( \boldsymbol{C}_{d} \right)$ over $d \in \mathcal{D}$. Then $d^*$ is universally optimal for the direct effects over $\mathcal{D}$.
\label{lemma1}
\end{lemma}

\begin{proof}
The proof is on similar lines as that of \cite{14}.
\end{proof}

\begin{lemma}
Let a design $d^* \in \mathcal{D}$, where $\mathcal{D}$ is a subclass of designs, be such that for $d \in \mathcal{D}$, the corresponding information matrix $\boldsymbol{C}_d \in \mathcal{B}_{gt}$; for any $d \in \mathcal{D}$ and $\boldsymbol{C}_{d} \neq \zero_{gt \times gt}$, there exists scalars $b_{d1}, \cdots, b_{d(gt)!} \geq 0$ satisfying $\boldsymbol{C}_{d^*} = \sum_{\kappa=1}^{(gt)!} b_{d\kappa} \boldsymbol{P}_{\kappa} \boldsymbol{C}_{d} \boldsymbol{P}^{'}_{\kappa}$, where $g>1$ and $\boldsymbol{P}_1$, $\cdots$, $\boldsymbol{P}_{(gt)!}$ are all possible distinct $gt \times gt$ permutation matrices. Then the design $d^*$ is universally optimal for the direct effects over $\mathcal{D}$, if $d^*$ maximizes $\text{\rmfamily\upshape tr} \left( \boldsymbol{C}_{d} \right)$ over $d \in \mathcal{D}$.
\label{lemma2}
\end{lemma}

\begin{proof}
The proof is similar to that in \cite{31}.
\end{proof}

Now, we consider designs with $p=t$. For the univariate response case with correlated error terms, \cite{17} proved that a design given as an orthogonal array of type $I$ and strength $2$, \\$OA_{I} \left( n=\lambda t \left(t-1 \right), p, t, 2 \right)$, where $3 \leq p \leq t$ and $\lambda$ is a positive integer, is universally optimal for the direct effects over the class of binary designs. Next, we check whether the universal optimality of $OA_{I} \left( n=\lambda t \left(t-1 \right), p=t, t, 2 \right)$, where $t \geq 3$ and $\lambda$ is a positive integer, holds for the multivariate response setup. Let us define the class of competing designs as
\begin{equation*}
\mathcal{D} = \{ d \in \Omega_{t,n,p=t} : d \text{ is a binary design and } t \geq 3   \}.
\end{equation*}
It should be noted that $\mathcal{D}$ is a subclass of designs which are uniform on subjects.

\begin{lemma}
Let $d^* \in \Omega_{t,n=\lambda t \left(t-1 \right),p=t}$ be a design given by $OA_{I} \left( n=\lambda t \left(t-1 \right), \right.$\\$\left.p=t, t, 2 \right)$, where $\lambda$ is a positive integer and $t \geq 3$. Then for $g > 1$, corresponding to design $d^*$, the information matrix for the direct effects can be expressed as 
\begin{equation}
\boldsymbol{C}_{d^*} =\oplus_{k=1}^{g} \sigma_k^{-2} \left(   \text{\rmfamily\upshape det} \left( \boldsymbol{Q}_k \right) / q_{(k)22} \right) \hatmat_t,
\label{le2c-1}
\end{equation}
where $\boldsymbol{Q}_k = \frac{n}{t-1} \begin{bmatrix}
q_{(k)11} & q_{(k)12}\\
q_{(k)12} & q_{(k)22}
\end{bmatrix}$, $\text{\rmfamily\upshape det} \left( \boldsymbol{Q}_k \right) \neq 0$, $q_{(k)11} = \text{\rmfamily\upshape tr} \left(  \boldsymbol{T}^{'}_{d^*1} \boldsymbol{V}_k^{*} \boldsymbol{T}_{d^*1}\right)$, $q_{(k)12} = \text{\rmfamily\upshape tr} \left(  \boldsymbol{T}^{'}_{d^*1} \boldsymbol{V}_k^{*} \boldsymbol{\psi} \boldsymbol{T}_{d^*1}\right)$ and $q_{(k)22} = \text{\rmfamily\upshape tr} \left(  \boldsymbol{T}^{'}_{d^*1} \boldsymbol{\psi}^{'} \boldsymbol{V}_k^{*} \boldsymbol{\psi} \boldsymbol{T}_{d^*1}\right) -  \frac{\left(\boldsymbol{V}_k^{*} \right)_{1,1}}{t}$, $\hatmat_t = \identity_t - \frac{1}{t} \vecone_{t} \vecone_{t}^{'}$, $\boldsymbol{\psi} =  
\begin{bmatrix}
\zero^{'}_{p-1 \times 1} & 0\\
\identity_{p-1} & \zero_{p-1 \times 1}
\end{bmatrix}$, $\boldsymbol{V}_k^* = \boldsymbol{V}_k^{-1} - \left( \vecone_p^{'} \boldsymbol{V}_k^{-1} \vecone_p \right)^{-1} \boldsymbol{V}_k^{-1} \vecone_{p} \vecone_{p}^{'} \boldsymbol{V}_k^{-1}$, and $\left(\boldsymbol{V}_k^{*} \right)_{1,1}$ is the element corresponding to first row and first column of the matrix $\boldsymbol{V}_k^{*}$.
\label{le2c}
\end{lemma}

\begin{proof}
Here $d^* \in \Omega_{t,n=\lambda t \left(t-1 \right),p=t}$ is a design given by $OA_{I} \left( n=\lambda t \left(t-1 \right), p=t, t, 2 \right)$, where $\lambda$ is a positive integer and $t \geq 3$. From \cite{23} and \cite{1}, we get that for $g=1$, the information matrix corresponding to $d^*$ for the direct effects is given as
\begin{equation*}
\boldsymbol{C}_{d^*} =\boldsymbol{T}^{'}_d  \boldsymbol{\Sigma}_1^{-1/2} \text{\rmfamily\upshape pr}^{\perp} \left( \boldsymbol{\Sigma}_1^{-1/2} 
\begin{bmatrix}
\boldsymbol{X_1} &  \boldsymbol{F}_d
\end{bmatrix}  
\right) \boldsymbol{\Sigma}_1^{-1/2} \boldsymbol{T}_d = \left(   \text{\rmfamily\upshape det} \left( \boldsymbol{Q}_1 \right) / q_{(1)22} \right) \hatmat_t,
\end{equation*}
where $\boldsymbol{Q}_1 = \frac{n}{t-1} \begin{bmatrix}
q_{(1)11} & q_{(1)12}\\
q_{(1)12} & q_{(1)22}
\end{bmatrix}$, $\text{\rmfamily\upshape det} \left( \boldsymbol{Q}_1 \right) \neq 0$, $q_{(1)11} = \text{\rmfamily\upshape tr} \left(  \boldsymbol{T}^{'}_{d^*1} \boldsymbol{V}_1^{*} \boldsymbol{T}_{d^*1}\right)$, $q_{(1)12} = \text{\rmfamily\upshape tr} \left(  \boldsymbol{T}^{'}_{d^*1} \boldsymbol{V}_1^{*} \boldsymbol{\psi} \boldsymbol{T}_{d^*1}\right)$ and $q_{(1)22} = \text{\rmfamily\upshape tr} \left(  \boldsymbol{T}^{'}_{d^*1} \boldsymbol{\psi}^{'} \boldsymbol{V}_1^{*} \boldsymbol{\psi} \boldsymbol{T}_{d^*1}\right) -  \frac{\left(\boldsymbol{V}_1^{*} \right)_{1,1}}{t}$, $\hatmat_t = \identity_t - \frac{1}{t} \vecone_{t} \vecone_{t}^{'}$, $\boldsymbol{\psi} =  
\begin{bmatrix}
\zero^{'}_{p-1 \times 1} & 0\\
\identity_{p-1} & \zero_{p-1 \times 1}
\end{bmatrix}$, $\boldsymbol{V}_1^{*} = \boldsymbol{V}_1^{-1} - \left( \vecone_p^{'} \boldsymbol{V}_1^{-1} \vecone_p \right)^{-1} \boldsymbol{V}_1^{-1} \vecone_{p} \vecone_{p}^{'} \boldsymbol{V}_1^{-1}$, and $\left(\boldsymbol{V}_1^{*} \right)_{1,1}$ is the element corresponding to first row and first column of the matrix $\boldsymbol{V}_1^{*}$. Here $\boldsymbol{V}_1$ is a known positive definite and symmetric matrix.\par
For the multivariate response setup, we have $\boldsymbol{V}_k$ to be a known positive definite and symmetric matrix. Using the expression of the information matrix from Lemma~\ref{thm4c}, we get
\begin{equation*}
\boldsymbol{C}_{d^*} =\oplus_{k=1}^{g} \sigma_k^{-2} \left(   \text{\rmfamily\upshape det} \left( \boldsymbol{Q}_k \right) / q_{(k)22} \right) \hatmat_t,
\end{equation*}
where $\boldsymbol{Q}_k = \frac{n}{t-1} \begin{bmatrix}
q_{(k)11} & q_{(k)12}\\
q_{(k)12} & q_{(k)22}
\end{bmatrix}$, $\text{\rmfamily\upshape det} \left( \boldsymbol{Q}_k \right) \neq 0$, $q_{(k)11} = \text{\rmfamily\upshape tr} \left(  \boldsymbol{T}^{'}_{d^*1} \boldsymbol{V}_k^{*} \boldsymbol{T}_{d^*1}\right)$, $q_{(k)12} = \text{\rmfamily\upshape tr} \left(  \boldsymbol{T}^{'}_{d^*1} \boldsymbol{V}_k^{*} \boldsymbol{\psi} \boldsymbol{T}_{d^*1}\right)$ and $q_{(k)22} = \text{\rmfamily\upshape tr} \left(  \boldsymbol{T}^{'}_{d^*1} \boldsymbol{\psi}^{'} \boldsymbol{V}_k^{*} \boldsymbol{\psi} \boldsymbol{T}_{d^*1}\right) -  \frac{\left(\boldsymbol{V}_k^{*} \right)_{1,1}}{t}$, $\boldsymbol{V}_k^* = \boldsymbol{V}_k^{-1} - \left( \vecone_p^{'} \boldsymbol{V}_k^{-1} \vecone_p \right)^{-1} \boldsymbol{V}_k^{-1} \vecone_{p} \vecone_{p}^{'} \boldsymbol{V}_k^{-1}$, and $\left(\boldsymbol{V}_k^{*} \right)_{1,1}$ is the element corresponding to first row and first column of the matrix $\boldsymbol{V}_k^{*}$.
\end{proof}

\begin{remark}
Let $d^* \in \Omega_{t,n=\lambda t \left(t-1 \right),p=t}$ be a design given by $OA_{I} \left( n=\lambda t  \left(t-1 \right), p=t, t, 2 \right)$, where $\lambda$ is a positive integer and $t \geq 3$. Then for $g > 1$, corresponding to the direct effects, the information matrix $\boldsymbol{C}_{d^*}$ is not completely symmetric.
\label{re2c}
\end{remark}

\begin{proof}
From Lemma~\ref{le2c}, we know that for $g>1$,
\begin{equation*}
\boldsymbol{C}_{d^*} =\oplus_{k=1}^{g} \sigma_k^{-2} \left(   \text{\rmfamily\upshape det} \left( \boldsymbol{Q}_k \right) / q_{(k)22} \right) \hatmat_t,
\end{equation*}
where $\boldsymbol{Q}_k = \frac{n}{t-1} \begin{bmatrix}
q_{(k)11} & q_{(k)12}\\
q_{(k)12} & q_{(k)22}
\end{bmatrix}$, $\text{\rmfamily\upshape det} \left( \boldsymbol{Q}_k \right) \neq 0$, $q_{(k)11} = \text{\rmfamily\upshape tr} \left(  \boldsymbol{T}^{'}_{d^*1} \boldsymbol{V}_k^{*} \boldsymbol{T}_{d^*1}\right)$, $q_{(k)12} = \text{\rmfamily\upshape tr} \left(  \boldsymbol{T}^{'}_{d^*1} \boldsymbol{V}_k^{*} \boldsymbol{\psi} \boldsymbol{T}_{d^*1}\right)$ and $q_{(k)22} = \text{\rmfamily\upshape tr} \left(  \boldsymbol{T}^{'}_{d^*1} \boldsymbol{\psi}^{'} \boldsymbol{V}_k^{*} \boldsymbol{\psi} \boldsymbol{T}_{d^*1}\right) -  \frac{\left(\boldsymbol{V}_k^{*} \right)_{1,1}}{t}$, $\hatmat_t = \identity_t - \frac{1}{t} \vecone_{t} \vecone_{t}^{'}$, $\boldsymbol{\psi} =  
\begin{bmatrix}
\zero^{'}_{p-1 \times 1} & 0\\
\identity_{p-1} & \zero_{p-1 \times 1}
\end{bmatrix}$, $\boldsymbol{V}_k^{*} = \boldsymbol{V}_k^{-1} - \left( \vecone_p^{'} \boldsymbol{V}_k^{-1} \vecone_p \right)^{-1} \boldsymbol{V}_k^{-1} \vecone_{p} \vecone_{p}^{'} \boldsymbol{V}_k^{-1}$, and $\left(\boldsymbol{V}_k^{*} \right)_{1,1}$ is the element corresponding to first row and first column of the matrix $\boldsymbol{V}_k^{*}$.\par
Since $\sigma_k^{-2}>0$, the necessary condition for the matrix $\boldsymbol{C}_{d^*}$ to be completely symmetric is that all the off-diagonal elements of $\left(   \text{\rmfamily\upshape det} \left( \boldsymbol{Q}_k \right) / q_{(k)22} \right) \hatmat_t$ are $0$. Since we know that off-diagonal elements of $\hatmat_t$ are $-1/t$ and for $t \geq 3$, $\text{\rmfamily\upshape det} \left( \boldsymbol{Q}_k \right) \neq 0$, the off-diagonal elements of $\left(   \text{\rmfamily\upshape det} \left( \boldsymbol{Q}_k \right)/ q_{(k)22} \right) \hatmat_t$ are non-zero. Thus for $g>1$ and $t \geq 3$, $\boldsymbol{C}_{d^*}$ is not completely symmetric.
\end{proof}

From Remark~\ref{re2c}, it is clear that we cannot use Lemma~\ref{lemma1}. Since $\left(   \text{\rmfamily\upshape det} \left( \boldsymbol{Q}_k \right) /  q_{(k)22} \right) \hatmat_t$ is a completely symmetric matrix, in order to apply Lemma~\ref{lemma2}, we require that for any $d \in \mathcal{D}$ and $\boldsymbol{C}_{d(k)} \neq \zero_{t \times t}$, $\sigma_k^{-2} \text{\rmfamily\upshape tr} \left( \boldsymbol{C}_{d^*(k)} \right) / \left[ t! \times \text{\rmfamily\upshape tr} \left( \boldsymbol{C}_{d(k)} \right) \right]$ is same for all $k$. Since $\boldsymbol{C}_{d(k)}$ depends on $k$ through $\boldsymbol{V}_k$, and $\sigma_k^2$ and $\boldsymbol{V}_k$ may vary according to $k$, this condition does not hold true. So, Lemma~\ref{lemma2} is not applicable. Thus we resort to identify $A$-, $D$- and $E$-optimal design.\par
Note that we have $gt$ direct effects. From \cite{1}, we have that $\boldsymbol{C}_{d(k)} \vecone_t = \zero_{t \times 1}$. So from Lemma~\ref{thm4c}, it is clear that for $g>1$, $\text{\rmfamily\upshape rank} \left( \boldsymbol{C}_d \right) \leq g(t-1)$. Thus all contrasts are not estimable. Hence, here we only consider designs such that for each variable, $t-1$ independent orthonormal contrasts of the direct effects are estimable.\par
To combine these $ t-1 $ contrasts for all variables, we introduce a $\left(t-1\right) \times t$ matrix $\boldsymbol{L}_k$, where $\text{\rmfamily\upshape rank} \left( \boldsymbol{L}_k \right) = t-1$, $ \boldsymbol{L}_k \vecone_{t} = \zero_{t \times 1} $ and $ \boldsymbol{L}_k \boldsymbol{L}_k^{'} = \identity_{t-1}$. So, $\boldsymbol{L}_k \boldsymbol{\tau}_k$ is the vector of $t-1$ independent orthonormal contrasts of the direct effects corresponding to the $k^{th}$ response variable. Let us denote $\mathbb{D} \left( \boldsymbol{L} \hat{\boldsymbol{\tau}} \right)$ by $\boldsymbol{G}_d$, where $\boldsymbol{L} \hat{\boldsymbol{\tau}}$ is the best linear unbiased estimator of $\boldsymbol{L} {\boldsymbol{\tau}}$ and
\begin{equation}
\boldsymbol{L} = \boldsymbol{L}_1  \oplus \cdots \oplus \boldsymbol{L}_g = \oplus_{k=1}^{g} \boldsymbol{L}_k.
\label{s2}
\end{equation}
Then a design $d^* \in \mathcal{D}$ is $A$-, $D$- and $E$-optimal for the direct effects over $\mathcal{D}$ under model \eqref{unio}, for $g \geq 1$, if $d^*$ minimizes $\text{\rmfamily\upshape tr} \left(  \boldsymbol{G}_d \right)$, $\text{\rmfamily\upshape det}  \left(\boldsymbol{G}_d \right)$ and the maximum eigenvalue of $\boldsymbol{G}_d$, respectively, over $d \in \mathcal{D}$.\par
From Lemma~\ref{thm4c}, we get that for $d \in \Omega_{t,n,p}$,
\begin{equation}
\mathbb{D} \left( \boldsymbol{L} 
\begin{bmatrix}
\hat{\boldsymbol{\tau}}^{'}_1 &
\hat{\boldsymbol{\tau}}^{'}_2 &
\cdots &
 \hat{\boldsymbol{\tau}}^{'}_g
\end{bmatrix}^{'} \right) = \boldsymbol{G}_d = \sigma_1^2 \boldsymbol{L}_1 \boldsymbol{C}_{d(1)}^{+} \boldsymbol{L}^{'}_1 \oplus  \cdots \oplus \sigma_g^2 \boldsymbol{L}_g \boldsymbol{C}_{d(g)}^{+} \boldsymbol{L}^{'}_g,
\label{s3new}
\end{equation} 
where $\boldsymbol{C}_{d(k)}^{+}$ is the Moore-Penrose inverse of the matrix $\boldsymbol{C}_{d(k)}$.\par
Minimizing $\sum_{k=1}^{g} \text{\rmfamily\upshape tr} \left( \sigma^2_k \boldsymbol{L}_k \boldsymbol{C}_{d(k)}^{+} \boldsymbol{L}^{'}_k \right) $, $\sum_{k=1}^{g} log \left( \text{\rmfamily\upshape det} \left( \sigma_k^2 \boldsymbol{L}_k \boldsymbol{C}_{d(k)}^{+} \boldsymbol{L}^{'}_k \right) \right)$ and the maximum eigenvalue of $\oplus_{k=1}^{g} \sigma^2_k \boldsymbol{L}_k \boldsymbol{C}_{d(k)}^{+} \boldsymbol{L}^{'}_k$ we can find $A$-, $D$- and $E$-optimal design, respectively.\par
Let $ \zeta_{d(k)1} \leq  \cdots \leq \zeta_{d(k)\left(t-1 \right)}$ be the positive eigenvalues of $ \boldsymbol{C}_{d(k)}$ and $\eta_{d(k)1} \leq  \cdots \leq \eta_{d(k)\left(t-1 \right)}$ be the eigenvalues of $ \sigma_k^2 \boldsymbol{L}_k  \boldsymbol{C}^{+}_{d(k)} \boldsymbol{L}^{'}_k $. Then following a similar approach as given in \cite{6}, we get that all the nonzero eigenvalues of $\boldsymbol{C}^{+}_{d(k)}$ are the eigenvalues of $\boldsymbol{L}_k \boldsymbol{C}^{+}_{d(k)}  \boldsymbol{L}_k^{'}$. Thus the eigenvalues of $\sigma_k^2 \boldsymbol{L}_k  \boldsymbol{C}^{+}_{d(k)} \boldsymbol{L}^{'}_k$ are
\begin{equation}
\eta_{d(k)1} = \frac{\sigma_k^2}{\zeta_{d(k)\left( t-1 \right)}},
\cdots,
\eta_{d(k)(t-1)} =  \frac{\sigma_k^2}{\zeta_{d(k)1}}.
\label{s8new}
\end{equation}
Using the above eigenvalue equations from \eqref{s8new}, we get
\begin{equation*}
\begin{split}
&\sum_{k=1}^{g} \text{\rmfamily\upshape tr} \left(  \sigma_k^2 \boldsymbol{L}_k \boldsymbol{C}_{d(k)}^{+} \boldsymbol{L}^{'}_k \right) = \sum_{k=1}^{g} \sum_{\omega=1}^{t-1} \eta_{d(k)\omega} = \sum_{k=1}^{g} \sum_{\omega=1}^{t-1} \frac{\sigma_k^2}{\zeta_{d(k)\omega}},\\
&\sum_{k=1}^{g} log \left(  \text{\rmfamily\upshape det} \left( \sigma_k^2 \boldsymbol{L}_k \boldsymbol{C}_{d(k)}^{+} \boldsymbol{L}^{'}_k \right) \right) =  \sum_{k=1}^{g} \sum_{\omega=1}^{t-1} log \left( \eta_{d(k) \omega} \right) = \sum_{k=1}^{g} \sum_{\omega=1}^{t-1} log \left( \frac{\sigma_k^2}{\zeta_{d(k)\omega}} \right), \text{ and}\\
&{\small{\text{The maximum eigenvalue of } \oplus_{k=1}^{g} \sigma_k^2 \boldsymbol{L}_k \boldsymbol{C}_{d(k)}^{+} \boldsymbol{L}^{'}_k = \max_{1 \leq k \leq g}~\eta_{d(k)(t-1)} =  \max_{1 \leq k \leq g}~\frac{\sigma_k^2}{ \zeta_{d(k)1}}}}.
\end{split}
\end{equation*}
Thus a design $d^* \in \mathcal{D}$ is $A$-, $D$- and $E$-optimal for the direct effects over $\mathcal{D}$ if $d^*$ minimizes $\sum_{k=1}^{g} \sum_{\omega=1}^{t-1} \frac{\sigma_k^2}{\zeta_{d(k)\omega}}$, $\sum_{k=1}^{g} \sum_{\omega=1}^{t-1} log \left( \frac{\sigma_k^2}{\zeta_{d(k)\omega}} \right)$ and $\max_{1 \leq k \leq g}~\sigma_k^2 \zeta^{-1}_{d(k)1}$, respectively, over $d \in \mathcal{D}$.

\begin{theorem}
Let $\mathcal{D}$ be a class of binary designs with $p=t$ and $d^* \in \mathcal{D}$ be a design represented as $OA_{I} \left( n=\lambda t \left(t-1 \right), p=t, t, 2 \right)$, where $\lambda$ is a positive integer and $t \geq 3$. Then under the $g>1$ case, $d^*$ is $A$-, $D$- and $E$-optimal for the direct effects over $\mathcal{D}$.
\label{thm32c}
\end{theorem}

\begin{proof}
Here $d^* \in \mathcal{D}$ is a design represented as $OA_{I} \left( n=\lambda t \left(t-1 \right), p=t, t, 2 \right)$, where $\mathcal{D}$ is a class of binary designs with $p=t \geq 3$ and $\lambda$ is a positive integer. From \cite{17}, we know that for $g=1$ setup, $d^*$ is universally optimal for the direct effects over $\mathcal{D}$. Since universal optimality implies $A$-, $D$- and $E$-optimality, we get that for $g=1$ and $t \geq 3$, $d^*$ minimizes $\sum_{\omega=1}^{t-1} \frac{\sigma_1^2}{\zeta_{d(1)\omega}}$, $ \sum_{\omega=1}^{t-1} log \left( \frac{\sigma_1^2}{\zeta_{d(1)\omega}} \right)$ and $\sigma_1^2\zeta^{-1}_{d(1)1}$ over $d \in \mathcal{D}$. Here $ \zeta_{d(1)1} \leq \cdots \leq \zeta_{d(1)\left(t-1 \right)}$ are the positive eigenvalues of $ \boldsymbol{C}_{d(1)}$, where $ \boldsymbol{C}_{d(1)}$ is as given in Lemma~\ref{thm4c}, and $\boldsymbol{V}_1$ is a known positive definite and symmetric $p \times p$ matrix.\par
Here, for $g>1$, we have $\boldsymbol{V}_k$ as a known positive definite and symmetric $p \times p$ matrix. So for $g>1$ and $t \geq 3$, for each $k$, $d^*$ minimizes $\sum_{\omega=1}^{t-1} \frac{\sigma_k^2}{\zeta_{d(k)\omega}}$, $ \sum_{\omega=1}^{t-1} log \left( \frac{\sigma_k^2}{\zeta_{d(k)\omega}} \right)$ and $\sigma_k^2 \zeta^{-1}_{d(k)1}$ over $d \in \mathcal{D}$. Hence for $g > 1$ and $t \geq 3$, $d^*$ minimizes {\small{$\sum_{k=1}^{g} \sum_{\omega=1}^{t-1} \frac{\sigma_k^2}{\zeta_{d(k)\omega}}$, $ \sum_{k=1}^{g} \sum_{\omega=1}^{t-1} log \left( \frac{\sigma_k^2}{\zeta_{d(k)\omega}} \right)$ and $\max_{1 \leq k \leq g}~\sigma_k^2 \zeta^{-1}_{d(k)1}$}} over $d \in \mathcal{D}$. Thus for $t \geq 3$, $d^*$ is $A$-, $D$- and $E$-optimal for the direct effects over $\mathcal{D}$ under the $g >1$ case.
\end{proof}

\section{Illustration}
In this section, we examine $A$-, $D$- and $E$-efficiency of a $3 \times 3$ crossover design using a gene profile study by \cite{22}. This design allocates treatment sequences $ACB$, $BAC$ and $CBA$ each to $6$ subjects, where $A$ and $B$ denote $10$ mg and $25$ mg dose of an oral drug, respectively, and $C$ denotes placebo. From \cite{25}, it is noted that in the gene profile, there exist pair of genes which are uncorrelated between themselves but may have within-response correlation.\par
Let $d_0$ represent the above $3\times 3$ design with treatment sequences $ACB$, $BAC$ and $CBA$. To find the $A$-, $D$- and $E$-efficiencies of this $d_0$, we assume a bivariate response setup, i.e., $g=2$. Here, we consider $\frac{\sigma_1^2}{\sigma_2^2} =2$ (any non-negative value other than $2$ could have been taken), $\boldsymbol{V}_1$ to have $AR(1)$ structure and $\boldsymbol{V}_2$ to have equi-correlated structure. Thus,
\begin{equation*}
\boldsymbol{V}_1 =
\begin{bmatrix}
1 & r_{(1)} & r_{(1)}^2\\
r_{(1)} & 1 & r_{(1)}\\
r_{(1)}^2 & r_{(1)} & 1
\end{bmatrix}
\quad 
\text{and} 
\quad
\boldsymbol{V}_2 =
\begin{bmatrix}
1 & r_{(1)} & r_{(1)}\\
r_{(1)} & 1 & r_{(1)}\\
r_{(1)} & r_{(1)} & 1
\end{bmatrix},
\end{equation*}
where $-0.5 < r_{(1)} < 1$. The constraint on the values of $r_{(1)}$ is to ensure positive definiteness of $\boldsymbol{V}_1$ and $\boldsymbol{V}_2$.\par
\begin{figure}[tb]
\centering
\begin{subfigure}{.5\textwidth}
  \includegraphics[height=\linewidth, width=1.0\linewidth]{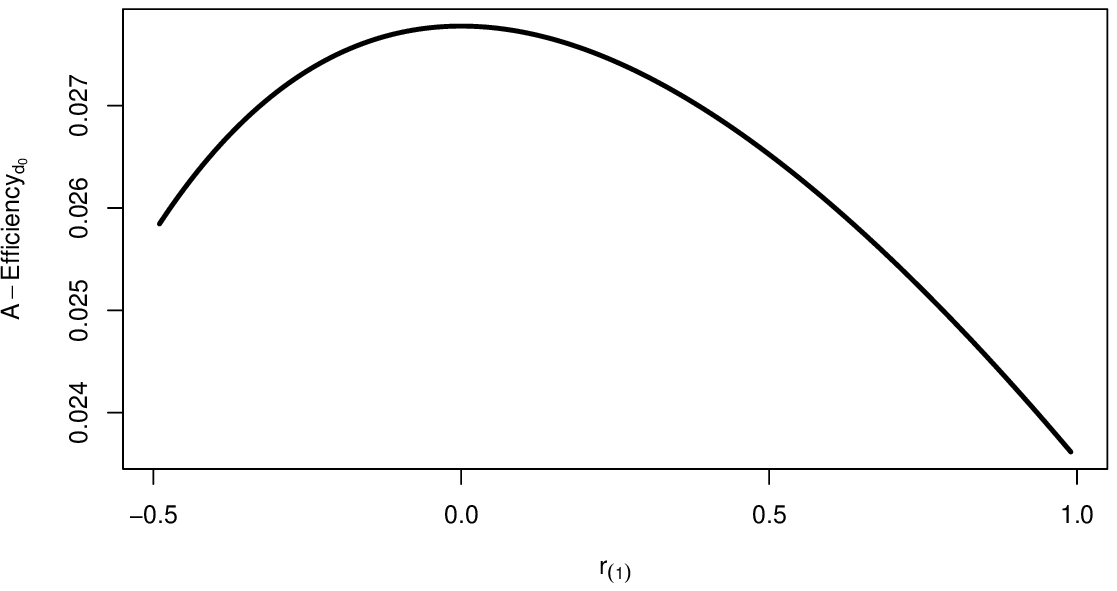}
  \caption{A-efficiency.}
  \label{A-Efficiency}
\end{subfigure}%
\begin{subfigure}{.5\textwidth}
\includegraphics[height=\linewidth, width=1.0\linewidth]{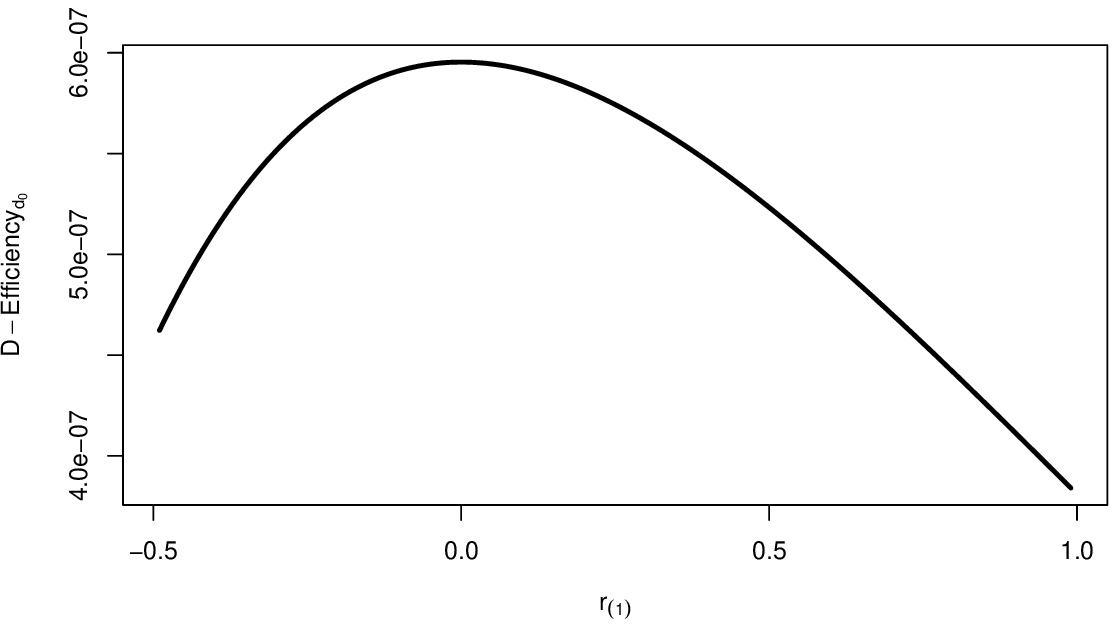}
  \caption{D-efficiency.}
  \label{D-Efficiency}
\end{subfigure}\\
\begin{subfigure}{.5\textwidth}
  \includegraphics[height=\linewidth, width=1.0\linewidth]{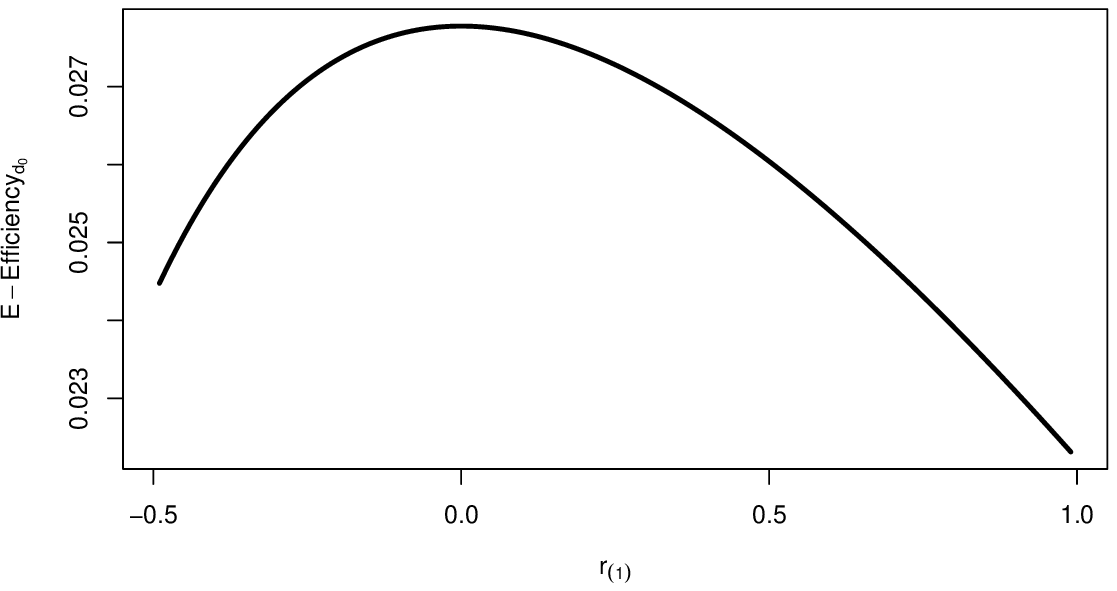}
  \caption{E-efficiency.}
  \label{E-Efficiency}
\end{subfigure}
\caption{Efficiency of $d_0$.}
\label{eff}
\end{figure}

From Theorem~\ref{thm32c}, we see that for $g>1$, a design $d^* \in \mathcal{D}$ represented by $OA_{I} \left( n=\lambda t \left(t-1 \right), p=t, t, 2 \right)$ is $A$-, $D$- and $E$-optimal for the direct effects over $\mathcal{D}$. We are interested in studying efficiency of the binary design $d_0$ for all permissible values of $r_{(1)}$. Suppose for a design $d_0$, the $A$-, $D$- and $E$-efficiency is given by
\begin{equation*}
\text{A-Efficiency}_{d_0} = \frac{\phi_A \left( \boldsymbol{C}_{d^*} \right)}{\phi_A \left( \boldsymbol{C}_{d_0} \right)}, \text{ }
\text{D-Efficiency}_{d_0} = \frac{\phi_D \left( \boldsymbol{C}_{d^*} \right)}{\phi_D \left( \boldsymbol{C}_{d_0} \right)}, \text{ and }
\end{equation*}
\begin{equation*}
\text{E-Efficiency}_{d_0} = \frac{\phi_E \left( \boldsymbol{C}_{d^*} \right)}{\phi_E \left( \boldsymbol{C}_{d_0} \right)},
\end{equation*}
where for $d \in \{  d_0, d^* \}$, $\phi_A \left( \boldsymbol{C}_{d} \right) = \sum_{k=1}^{g} \sum_{\omega=1}^{t-1} \frac{\sigma_k^2}{\zeta_{d(k)\omega}}$, $\phi_D \left( \boldsymbol{C}_{d} \right) =\prod_{k=1}^{g} \prod_{\omega=1}^{t-1} \frac{\sigma_k^2}{\zeta_{d(k)\omega}}$ and $\phi_E \left( \boldsymbol{C}_{d} \right) = \max_{1 \leq k \leq g}~\frac{\sigma_k^2}{ \zeta_{d(k)1}}$. It should be noted that minimization of $\sum_{k=1}^{g} \sum_{\omega=1}^{t-1} log \left( \frac{\sigma_k^2}{\zeta_{d(k)\omega}} \right)$ is equivalent to minimization of $ \prod_{k=1}^{g} \prod_{\omega=1}^{t-1} \frac{\sigma_k^2}{\zeta_{d(k)\omega}}$. Also, note that $\phi_A \left( \boldsymbol{C}_{d} \right)$, $\phi_D \left( \boldsymbol{C}_{d} \right)$ and $\phi_E \left( \boldsymbol{C}_{d} \right)$ depends on $r_{(1)}$ only through the eigenvalues, $\zeta_{d(k)\omega}$. The values of A-, D- and $\text{E-Efficiency}_{d_0}$ lie between $0$ and $1$, and a value closer to $1$ indicates that design $d$ is a nearly efficient design. From Figure~\ref{eff}, we see that $A$-, $D$- and $E$-efficiency of $d_0$ is at most $0.0278$, $5.9537 \times 10^{-7}$ and $0.0278$, respectively. Thus, implying that the experimenter would have obtained much higher efficiency if they had used an orthogonal array type design in the gene data example.

\section{Conclusion}
In this article, we considered a multivariate fixed effect model that accounts for within-response correlation. While the $g$ responses were assumed to be uncorrelated, the within-response correlation structure was allowed to vary between distinct responses. Corresponding to the direct effects, the information matrix deviated from the $g=1$ case under this error structure, particularly concerning the complete symmetricity. We demonstrated that an orthogonal array design of type $I$ with strength $2$ is $A$-, $D$-, and $E$-optimal for the direct effects over a class of binary designs with $p = t \geq 3$ in the $g > 1$ scenario. Under an uncorrelated but heteroscedastic error structure, it can be shown using similar techniques, if the universal optimality holds for a balanced uniform design over a subclass of designs for the direct effects (or carryover effects) in the univariate response case, then over the same subclass of designs it remains $A$-, $D$-, and $E$-optimal for the direct effects (or carryover effects) under the multivariate response setup.\par
Though in this article, we considered a fixed effect model, a mixed effects model with random subject effects may be considered. In \cite{12}, exact optimal designs for a univariate mixed effects model with random subject effects having independent and identical normal distribution are studied. This model may be extended to the multivariate mixed effects model by considering independent and identically distributed error terms following normal distribution. In this case, the responses will be uncorrelated between themselves, but the random subject effects will introduce correlation within responses, to be specific the within-response correlation structure will be equi-correlated. We can also consider the case where error terms have within-response correlation, but are uncorrelated between responses. In this scenario, the dispersion matrix of the observations on responses will have a complicated structure and hence it may be tedious to identify a theoretical exact optimal design. As a future direction, we intend to explore the $A$-, $D$- and $E$-optimality aspects in these scenarios for the multiple response case.

	\bibliographystyle{amsalpha}

\end{document}